\documentclass[preprint,aps,prd,superscriptaddress,preprintnumbers,amsmath,amssymb]{revtex4}

\newcommand{\fft}[2]{{\frac{#1}{#2}}}
\newcommand{\ft}[2]{{\textstyle\frac{#1}{#2}}}

\begin{document}
\preprint{MCTP-11-32}

\title{A holographic $c$-theorem for higher derivative gravity}

\author{James T.~Liu}
\email{jimliu@umich.edu}
\affiliation{Michigan Center for Theoretical Physics,
Randall Laboratory of Physics,
The University of Michigan,
Ann Arbor, MI 48109--1040, USA}

\author{Zhichen Zhao}
\email{zhichen@umich.edu}
\affiliation{Michigan Center for Theoretical Physics,
Randall Laboratory of Physics,
The University of Michigan,
Ann Arbor, MI 48109--1040, USA}

\begin{abstract}

We prove a holographic $c$-theorem for the $a$ central charge in AdS/CFT where the
bulk is described by a gravitational action built out of an arbitrary function $f(R^{ab}{}_{cd})$
of the Riemann tensor coupled to bulk matter.  This theorem holds provided a generalized
null energy condition involving both matter and higher curvature gravitational interactions
is satisfied.  As an example, we consider the case of a  curvature-squared action, and
find that generically the generalized null energy condition involves not just the bulk matter,
but also the sign of $R''$ where a prime denotes a radial derivative and where $R$ is the bulk
scalar curvature.

\end{abstract}

\maketitle

\section{Introduction}

The central charge $c$ is one of the most important quantities characterizing a two-dimensional
conformal theory.  It is related to the Weyl anomaly through
\begin{equation}
\langle T^\mu_\mu\rangle=-\fft{c}{12}R,
\end{equation}
and gives a measure of the number of degrees of freedom of the theory.  For flows in the
space of conformal field theories, the Zamolodchikov $c$-theorem \cite{Zamolodchikov:1986gt}
states that there exists
a $c$-function which is monotonically decreasing along flows from the UV to the IR, and
which is equal to the central charge at the fixed points of the flow.  The consequence of this
theorem, which is proven for any unitary quantum field theory in two dimensions (with
mild assumptions), is that UV degrees of freedom are removed as the theory flows to the IR.

There have been various attempts to generalize the Zamolodchikov $c$-theorem to higher
dimensions, where higher dimensional analogs of the two-dimensional central charge are
given by the coefficients of the Weyl anomaly.  In particular, in four dimensions, there are
two anomaly coefficients, $a$ and $c$, conventionally defined by
\begin{equation}
\langle T^\mu_\mu\rangle=\fft{c}{16\pi^2}C_{\mu\nu\rho\sigma}^2-\fft{a}{16\pi^2}E_4,
\end{equation}
where $E_4=R^2-4R_{\mu\nu}^2+R_{\mu\nu\rho\sigma}^2$ is the four-dimensional
Euler density.  The $a$-anomaly generalizes to any even dimension, and Cardy conjectured
that it is monotonically decreasing along flows to the IR \cite{Cardy:1988cwa}.
Support for such a higher-dimensional
$c$-theorem has been given in various situations
\cite{Osborn:1989td,Jack:1990eb,Anselmi:1997am,Anselmi:1997ys}
and in the context of $a$-maximization
\cite{Intriligator:2003jj,Kutasov:2003iy,Kutasov:2003ux,Barnes:2004jj}, and a recent
proof has been given in four dimensions in \cite{Komargodski:2011vj}.

Complementing the field theoretic proof of Cardy's conjecture, the advent of AdS/CFT
has opened up a new approach to the $c$-theorem based on holographic computations of
the Weyl anomaly \cite{Henningson:1998gx}.  Initially, a $c$-theorem has been proven for
holographic RG flows
described by Einstein gravity coupled to bulk matter satisfying the null energy condition
\cite{Alvarez:1998wr,Girardello:1998pd,Freedman:1999gp,Sahakian:1999bd}.
Since this corresponds to leading order in the $1/N$ expansion, the four-dimensional
central charges are equal in this case ($c=a$), and hence no distinction was made between
flows of $a$ and flows of $c$.  More recently, there have been several investigations
involving higher derivative gravity in the bulk, where $a$ and $c$ are no longer identified
\cite{Sinha:2010ai,Oliva:2010eb,Myers:2010xs,Sinha:2010pm,Myers:2010tj,Liu:2010xc}.

There is now convincing evidence that the holographic $c$-theorem holds for the $a$
central charge in cases where the bulk higher derivative gravity propagates fluctuations
with no higher than second order equations of motion.  This includes Gauss-Bonnet
and quasi-topological gravity
\cite{Sinha:2010ai,Oliva:2010eb,Myers:2010xs,Sinha:2010pm,Myers:2010tj}
as well as general Lovelock theories \cite{Liu:2010xc}.  Although such theories are
somewhat special, it was argued in \cite{Myers:2010tj} that the restriction to second order
equations of motion is necessary to ensure unitarity of the boundary field theory.  
Nevertheless, if we take an effective field theory approach to gravity, then perhaps this
restriction ought to be relaxed.  For example, in a stringy context, one would expect both
bulk and boundary theories to be well behaved, even though this may not be apparent
at any finite order in the low energy effective theory.

In \cite{Liu:2010xc}, we investigated the possibility of constructing a holographic $c$-theorem
for $f(R)$ gravity, and demonstrated that monoticity of the $a$ central charge will be ensured
by a combination of the null energy condition on the bulk matter and a condition related to
the higher derivative nature of the gravitational sector.  This second condition appears to be
a weaker version of the unitarity requirement of \cite{Myers:2010tj}.  In this letter, we
extend the investigation of \cite{Liu:2010xc} by constructing a general $a$-function which is
applicable to a large class of higher derivative gravity theories in the bulk coupled to matter.
As in the $f(R)$ case, monoticity is ensured by a combination of conditions on the bulk
matter and gravitational sectors of the theory.

\section{Defining the $a$-function}

Consider a general bulk action of the form
\begin{equation}
S=\fft1{2\kappa^2}\int d^{d+1}x\sqrt{-g}f(R^{ab}{}_{cd}) +S_{\rm matter}.
\label{action}
\end{equation}
Here, $a,b,\ldots=0,1,\ldots,d$ correspond to bulk indices, and below we will use
$\mu,\nu,\ldots=0,1,\ldots,d-1$ to denote boundary indices.
For simplicity, we take $f(R^{ab}{}_{cd})$ to be built out of the Riemann
tensor with raised and lowered indices as indicated.  In particular, it does not contain
any implicit metric factors nor derivatives of the Riemann tensor.  In this case, the Einstein
equation reads
\begin{equation}
G_{ab}\equiv F_{(a}{}^{cde}R_{b)cde}
-\ft12fg_{ab}+2\nabla^c\nabla^dF_{acbd}=\kappa^2T_{ab},
\label{eq:eins}
\end{equation}
where
\begin{equation}
F_{ab}{}^{cd}=\fft{\delta f(R^{ef}{}_{gh})}{\delta R^{ab}{}_{cd}}.
\end{equation}

Following \cite{Myers:2010tj,Liu:2010xc}, we consider the metric
\begin{equation}
ds^2=e^{2A(r)}(-dt^2+d\vec x_{d-1}^2)+dr^2,
\label{metric}
\end{equation}
and define the $a$-function
\begin{equation}
a(r)=\fft{2d\pi^{d/2}}{\kappa^2(d/2)!^2}\fft{F^{tr}{}_{tr}}{(A')^{d-1}}.
\label{eq:adef}
\end{equation}
As shown in \cite{Myers:2010tj,Liu:2010xc}, this function reproduces the holographic
$a$-anomaly \cite{Henningson:1998gx,Imbimbo:1999bj} when evaluated at an AdS
fixed point of the holographic flow.  To see this, consider, for example, the case of
Einstein gravity where $F^{tr}{}_{tr}=1/2$.  In this case, the $a$-function takes the form
\begin{equation}
a(r)=\fft{d\pi^{d/2}}{\kappa^2(d/2)!^2}\ell_{\rm eff}^{d-1},
\end{equation}
where $\ell_{\rm eff}=1/A'$ is the `effective' AdS radius.  Setting $d=4$ and taking
$\ell_{\rm eff}$ to be a constant AdS radius $L$ then gives the familiar holographic
expression \cite{Henningson:1998gx}
\begin{equation}
a=\fft{\pi^2}{\kappa_5^2}L^3=\fft{N^2}4,
\end{equation}
where we have used the AdS$_5$/CFT$_4$ dictionary $L^4=4\pi\alpha'^2g_sN$ and
$2\kappa_5^2=2\kappa_{10}^2/{\rm Vol}(S^5)=(2\pi)^7g_s^2\alpha'^4/\pi^3L^5$.

In the more general higher curvature case, the expression (\ref{eq:adef}) is somewhat
reminiscent of the Wald entropy formula \cite{Wald:1993nt,Iyer:1994ys,Iyer:1995kg}
\begin{equation}
S=\fft{4\pi}{\kappa^2}\int_\Sigma d^{d-1}x\sqrt{h}F^{tr}{}_{tr},
\end{equation}
where the integral is taken over the horizon of the black hole.  Although the bulk spacetimes
we are interested in are not necessarily that of black holes, we may nevertheless define
an entropy function \cite{Goldstein:2005rr,Cremades:2006ke}
\begin{equation}
\tilde C(r)=\fft{4\pi}{\kappa^2}e^{(d-1)A}F^{tr}{}_{tr},
\label{eq:ctfunc}
\end{equation}
corresponding to the metric (\ref{metric}).  While we are primarily focused on the
holographic $a$ anomaly, we will also comment on the behavior of $\tilde C(r)$ below.

\section{A holographic $c$-theorem and the null energy condition}

Given the $a$-function (\ref{eq:adef}), we now wish to demonstrate that $a'(r)\ge0$, or
at least understand the conditions for this to hold.  Motivated by the holographic
$c$-theorem in ordinary Einstein gravity
\cite{Alvarez:1998wr,Girardello:1998pd,Freedman:1999gp,Sahakian:1999bd},
which makes crucial use of the null energy condition, we focus on the combination
$G^t_t-G^r_r$, where the generalized Einstein tensor $G_{ab}$ is defined in (\ref{eq:eins}).
For the metric (\ref{metric}), this takes the form
\begin{equation}
G^t_t-G^r_r=2(d-1)(F^{tx}{}_{tx}R^{tx}{}_{tx}-F^{tr}{}_{tr}R^{tr}{}_{tr})+2\nabla_a\nabla^b
(F^{ta}{}_{tb}-F^{ra}{}_{rb}).
\end{equation}
Note that, because of the isometries of the metric,  we have taken $x$ to be an arbitrary
spatial direction (which we can take to be $x^1$).

Computing the Riemann tensor corresponding to the metric (\ref{metric}) is
straightforward, and gives
\begin{equation}
R^{\mu\nu}{}_{\rho\sigma}=-A'^2(\delta^\mu_\rho\delta^\nu_\sigma-\delta^\mu_\sigma
\delta^\nu_\rho),\qquad R^{\mu r}{}_{\nu r}=-(A''+A'^2)\delta^\mu_\nu.
\label{eq:Rcompute}
\end{equation}
The covariant derivatives acting on $F^{ab}{}_{cd}$ are somewhat more cumbersome
to evaluate.  We find
\begin{eqnarray}
\nabla_a\nabla^bF^{ta}{}_{tb}&=&
F^{tr}{}_{tr}{}''+(d-1)[A'(2F^{tr}{}_{tr}{}'-F^{tx}{}_{tx}{}')
+(A''+(d-1)A'^2)(F^{tr}{}_{tr}-F^{tx}{}_{tx})],\nonumber\\
\nabla_a\nabla^bF^{ra}{}_{rb}&=&dA'F^{tr}{}_{tr}{}'+d(d-1)A'^2(F^{tr}{}_{tr}-F^{tx}{}_{tx}).
\end{eqnarray}
Combining the above, we obtain
\begin{equation}
G^t_t-G^r_r=2(d-1)A''F^{tr}{}_{tr}-2A'F^{tr}{}_{tr}{}'+2F^{tr}{}_{tr}{}''
+2(d-1)[A'(F^{tr}{}_{tr}-F^{tx}{}_{tx})]'.
\label{eq:gttgrr}
\end{equation}

We are now in a position to examine flows of the $a$-function.  Taking a
radial derivative of (\ref{eq:adef}) and substituting in (\ref{eq:gttgrr}) yields
\begin{equation}
a'(r)=\fft{d\pi^{d/2}}{\kappa^2(d/2)!^2}\fft{-(G^t_t-G^r_r)+2[(d-1)A'(F^{tr}{}_{tr}-F^{tx}{}_{tx})
+F^{tr}{}_{tr}{}']'}{(A')^d}.
\label{eq:aprime}
\end{equation}
Assuming $d$ is even, the sign of $a'(r)$ is then given by the sign of the numerator above.
By imposing the null energy condition on bulk matter, $-(T^t_t-T^r_r)\ge0$, and using the
Einstein equation, we see that the first term in the numerator is indeed non-negative.
However, there does not appear to be any obvious constraint on the sign of the
second term.  What this demonstrates
is that the null energy condition by itself is no longer sufficient to guarantee monoticity of
the $a$-function in a higher derivative bulk theory of gravity.  Instead, what is required is
that the entire numerator is non-negative.  At the same time, however, there
appears to be a clear physical separation between the two terms in the numerator;
the first term is related to the bulk matter, while the second is related to the higher derivative
gravitational interactions.  As suggested in \cite{Liu:2010xc}, the sign of the latter term
may be connected to unitarity and ghost issues in the gravitational sector.  

One way to ensure that the $a$-function defined in (\ref{eq:adef}) is
monotonic increasing in flows to the UV is to impose separate conditions on the matter and
gravitational sectors: ($i$) the bulk matter must satisfy the null energy condition
$-(T^t_t-T^r_r)\ge0$, and ($ii$) the gravitational sector must satisfy $\Delta'\ge0$ where
\begin{equation}
\Delta\equiv (d-1)A'(F^{tr}{}_{tr}-F^{tx}{}_{tx})+F^{tr}{}_{tr}{}'.
\label{eq:Delta}
\end{equation}
Of course $\Delta=0$ for Einstein gravity.  Furthermore, for $f(R)$ gravity, the function
$F_{ab}{}^{cd}$ is simply
\begin{equation}
F^{ab}{}_{cd}=\ft12(\delta^a_c\delta^b_d-\delta^a_d\delta^b_c)F(R).
\end{equation}
Hence $\Delta=F'(R)/2$, and the condition $\Delta'\ge0$ is identical to the condition $F''\ge0$
obtained in \cite{Liu:2010xc}.  Note that this condition is entirely expressed in terms of
the scalar curvature $R$, and in particular does not involve explicit factors of the metric
scale factor $A$.

To further develop our understanding of the $\Delta'\ge0$ condition, we may examine a
general $R^2$ action of the form
\begin{equation}
f(R^{ab}{}_{cd})=R+\Lambda + \alpha_1 R^2 + \alpha_2 R_{ab} R^{ab}
+\alpha_3 R_{abcd}R^{abcd}.
\end{equation}
In this case, $F^{ab}{}_{cd}$ takes the form
\begin{equation}
F^{ab}{}_{cd}=
(1+2\alpha_1 R)\ft12(\delta^a_c\delta^b_d-\delta^a_d\delta^b_c)
+2\alpha_2\ft14(\delta^a_cR^b_d-\delta^a_dR^b_c-\delta^b_cR^a_d+\delta^b_dR^a_c)
+2\alpha_3R^{ab}{}_{cd}.
\end{equation}
Decomposing this into $\mu$ and $r$ components, and using (\ref{eq:Rcompute}), we obtain
\begin{equation}
\Delta=\fft{4d\alpha_1+(d+1)\alpha_2+4\alpha_3}{4d}R',
\label{eq:R2Delta}
\end{equation}
where $R=-d(2A''+(d+1)A'^2)$.  What is curious is that this is again written only in terms
of the curvature scalar $R$, even though $f(R^{ab}{}_{cd})$ involves the full Ricci tensor.

Note that $\Delta$ vanishes identically when $4d\alpha_1+(d+1)\alpha_2+4\alpha_3=0$,
as noted in \cite{Myers:2010tj}.  This encompasses the Gauss-Bonnet combination
$\{\alpha_1,\alpha_2,\alpha_3\}=\{1,-4,1\}$, but also allows for a two-parameter family of
$R^2$ theories that satisfy the $c$-theorem with only the null energy condition.

\section{DIscussion}

As we have seen, monoticity of flows of $a(r)$ for a general higher derivative bulk theory
follows from both the null energy condition and a gravitational sector condition $\Delta'\ge0$.
The latter condition is explicitly higher than second order in derivatives, and encodes the
content of the higher curvature terms in the bulk gravitational action.  As argued in
\cite{Myers:2010tj}, we would like to impose the physical requirement that the boundary
theory is unitary, both at fixed points and along the RG flow.  This requirement corresponds
to the constraint that the bulk metric fluctuations are second order in derivatives, and is
thus equivalent to demanding that $\Delta=0$.

While the restriction to bulk theories with only second order fluctuations makes physical
sense, we nevertheless believe additional information can be obtained even in the
presence of higher order fluctuations.  After all, we ought to view the bulk theory as an
effective gravity theory, in which case any breakdown in unitarity could potentially be
relegated to energy scales above those of interest.  For example, in the case of $R^2$
gravity, we have found a non-trivial $\Delta$ given in (\ref{eq:R2Delta}).  However, at
linearized order, a field redefinition $g_{ab}\to g_{ab}+\lambda_1R_{ab}
+\lambda_2g_{ab}R$ can be used to put $f(R^{ab}{}_{cd})$ into the Gauss-Bonnet form
\begin{equation}
f(R^{ab}{}_{cd})=R+\Lambda+\alpha_3(R_{abcd}R^{abcd}-4R_{ab}R^{ab}+R^2)+\cdots.
\end{equation}
In this frame, the linearized action is second order in derivatives, and $\Delta$ vanishes
identically.  Of course, what we have done is pushed the higher derivative interactions to
higher orders.  Nevertheless, this demonstrates that in some cases a theory with non-trivial
$\Delta$ (and hence explicit higher derivative terms) may be equivalent up to any finite
order in the higher curvature expansion to a theory with $\Delta=0$.

In fact, the issue of field redefinitions is somewhat more involved for the matter coupled
gravity system in the bulk.  For $R^2$ gravity at linearized order, one expects that only
$\alpha_3$ is physical, and yet both $\alpha_1$ and $\alpha_2$ show up in the expression
(\ref{eq:R2Delta}) for $\Delta$.  While $\Delta$ itself is not a direct physical observable,
its behavior along radial flows does have an impact on the monoticity of $a(r)$.  Thus its
dependence on $\alpha_1$ and $\alpha_2$ is perhaps somewhat unexpected.  We
believe that the resolution to this apparent paradox is that the above field redefinition
of the metric necessarily mixes the gravitational and matter sectors of the original action
(\ref{action}), so that in particular $S_{\mathrm{matter}}$ will now depend on both
curvature and actual matter fields.  This suggests that there is in fact no sharp distinction
between the gravity and matter sectors of the bulk theory, and that unitarity of the gravity
theory cannot be completely disentangled from unitarity of the bulk matter.

With the above in mind, we note that the
Einstein equation (\ref{eq:eins}) may be used to rewrite (\ref{eq:aprime}) as
\begin{equation}
a'(r)=\fft{d\pi^{d/2}}{\kappa^2(d/2)!^2}\fft{-\kappa^2(T^t_t-T^r_r)+2\Delta'}{(A')^d}.
\end{equation}
In this case, a weaker form of the $c$-theorem may be obtained by demanding only
that the numerator is non-negative
\begin{equation}
-\kappa^2(T^t_t-T^r_r)+2\Delta'\ge0.
\end{equation}
We believe this may be viewed as a generalized null energy condition that takes both
matter and gravity into account.  It would be interesting to see if this form of a generalized
null energy condition can be made more precise.  In particular, we believe this ought to be
directly related to the unitarity of dilaton scattering in the field theoretic proof of the $c$-theorem
presented in \cite{Komargodski:2011vj}.

In addition to the holographic Weyl anomaly and the $a$-function, it is also possible
to define an entropy function, (\ref{eq:ctfunc}), based on the Wald entropy formula.  Although
both $a(r)$ and $\tilde C(r)$ depend linearly on $F^{tr}{}_{tr}$, they differ in their dependence
on metric factors: $a(r)$ depends on the effective radius $\ell_{\rm eff}^{d-1}=1/(A')^{d-1}$,
while $\tilde C(r)$ depends on the horizon volume $e^{(d-1)A}$.  As a result, the proof of the
second law of black hole thermodynamics involving $\tilde C(r)$ differs from that of the
holographic $c$-theorem given above. 

It is in fact instructive to contrast flows of $a(r)$ with flows of $\tilde C(r)$.  For the latter, we
consider an affinely parameterized null congruence given by the tangent vector
$k^\mu\partial_\mu=d/d\lambda$ and define
\begin{equation}
\tilde\theta=\fft{d\log\tilde C}{d\lambda}=\theta+k^\mu\partial_\mu\log F^{tr}{}_{tr},
\end{equation}
where $\theta$ is the expansion of the null congruence.  In particular, for the
metric given in (\ref{metric}), we take $d/d\lambda=-e^{-2A}\partial_t+e^{-A}\partial_r$,
in which case
\begin{equation}
\tilde\theta=e^{-A}\left[(d-1)A'+\fft{F^{tr}{}_{tr}{}'}{F^{tr}{}_{tr}}\right].
\end{equation}
Taking a further radial derivative gives
\begin{equation}
\tilde\theta'=-e^{-A}\left[(d-1)A'^2+\left(\fft{F^{tr}{}_{tr}{}'}{F^{tr}{}_{tr}}\right)^2
-\fft{(d-1)A''F^{tr}{}_{tr}-A'F^{tr}{}_{tr}{}'+F^{tr}{}_{tr}{}''}{F^{tr}{}_{tr}}\right].
\end{equation}
We now substitute in the combination $G^t_t-G^r_r$ given in (\ref{eq:gttgrr}) to obtain
\begin{eqnarray}
\tilde\theta'&=&-e^{-A}\left[(d-1)A'^2+\left(\fft{F^{tr}{}_{tr}{}'}{F^{tr}{}_{tr}}\right)^2
+\fft{-\kappa^2(T^t_t-T^r_r)+2(d-1)[A'(F^{tr}{}_{tr}-F^{tx}{}_{tx})]'}{2F^{tr}{}_{tr}}\right].
\end{eqnarray}
Since the first two terms in the square brackets are non-negative, monoticity of $\tilde\theta'$
will be ensured so long as the third term is also non-negative.  This again involves something
like a generalized null energy condition
\begin{equation}
-\kappa^2(T^t_t-T^r_r)+2\tilde\Delta'\ge0,
\end{equation}
however with a different $\tilde\Delta=(d-1)A'(F^{tr}{}_{tr}-F^{tx}{}_{tx})$ from that given
in (\ref{eq:Delta}).  Note that $\tilde\Delta$ vanishes for $f(R)$ gravity
\cite{Cremades:2006ke,Jacobson:1995uq}, as in this case $F^{tr}{}_{tr}=F^{tx}{}_{tx}=F(R)/2$.
However, for $R^2$ gravity, we have
\begin{equation}
\tilde\Delta=-\fft{(d-1)((d-1)\alpha_2+4\alpha_3)}4(A'^2)'.
\end{equation}
In particular, this does not vanish for the Gauss-Bonnet combination.  This leads to
the curious observation that the null energy condition is sufficient to prove monoticity
of the $a$-function, yet does not appear to be sufficient for ensuing monoticity of the entropy 
function.

Finally, recent developments suggest that the $a$-anomaly is closely related to
the holographic entanglement entropy of the boundary CFT
\cite{Myers:2010xs,Myers:2010tj,Dowker:2010nq,deBoer:2011wk,Hung:2011xb,Casini:2011kv}.
This connection suggests that the holographic $c$-theorem is a universal means of
capturing the effective number of degrees of freedom along renormalization group flows of
the boundary theory.

\acknowledgments
Some of the ideas for handling the gravitational higher derivative corrections
were inspired by discussions at the Great Lakes Strings 2011 Conference
at the University of Chicago.
This work was supported in part by the US Department of Energy under grant
DE-FG02-95ER40899.


\end{document}